\documentclass[preprint,amsmath,amssymb,aps,nofootinbib]{revtex4}
\usepackage{graphicx}
\usepackage[utf8]{inputenc}
\usepackage{bm}
\usepackage{subfigure}
\usepackage{amsmath}

\newcommand{\pr}{Phys. Rev.\ }

\newcommand{\jpa}{J. Phys. A\ }
\newcommand{\jpb}{J. Phys. B\ }

\newcommand{\etal}{{\em et al. }}

\newcommand{\etals}{{\em et al.}}

\begin{document}

\title{Quantum correlations across two octaves from combined up and down conversion}

\author{Jingyan Li$^{1,2}$ and M.~K. Olsen$^{2,3}$}
\affiliation{\\
$^{1}$School of Electronic and Electrical Engineering, Wuhan Textile
University, Wuhan 430079, China\\
}
\affiliation{$^{2}$Quantum Science Otago and Dodd-Walls Centre for Photonic and Quantum
Technologies, Department of Physics, University of Otago, Dunedin, New
Zealand\\
}
\affiliation{$^{3}$School of Mathematics and Physics, University of Queensland, Brisbane,
Queensland 4072, Australia\\
}
\date{\today}

\begin{abstract}

We propose and analyse a cascaded optical parametric system which involves three interacting modes across two octaves of frequency difference. Our system, combining degenerate optical parametric oscillation (OPO) with second harmonic generation (SHG), promises to be a useful source of squeezed and entangled light at three differing frequencies. We show how changes in damping rates and the ratio of the two concurrent nonlinearities affect the quantum correlations in the output fields. We analyse the threshold behaviour, showing how the normal OPO threshold is changed by the addition of the SHG interactions. We also find that the inclusion of the OPO interaction removes the self-pulsing behaviour found in normal SHG. Finally, we show how the Einstein-Podolsky-Rosen correlations can be controlled by the injection of a coherent seed field at the lower frequency.    

\end{abstract}

\maketitle





\section{Introduction}
\label{sec:intro}

The theory of the interaction of light fields at one frequency with
nonlinear materials to produce fields at different frequencies goes back at
least to Armstrong \etal and their seminal work which included downconversion and second and third harmonic generation~\cite{Armstrong}. Since the publication of that work, the optical parametric oscillator (OPO) in both its degenerate and nondegenerate forms~\cite{Giordmaine,Nassau,Yariv} has become a standard workhorse for quantum optics and quantum information, especially with respect to the Einstein Podolsky Rosen paradox~\cite{RMPMargaret}. The related process of intracavity second harmonic generation (SHG) has also long been known to produce quantum states of the optical field~\cite{SHGPereira}. 

In the degenerate OPO, any entanglement will necessarily be across one octave, with the same being true of 
SHG~\cite{sumdiff,PingKoy}. In this work we combine these two processes in either a cascaded or concurrent manner, to produce entangled beams and states exhibiting EPR steering across two octaves of frequency difference.  Such a difference in frequencies has previously been predicted for a system which cascades two SHG processes to produce entangled outputs at three different frequencies, with both bipartite~\cite{4HG} and tripartite correlations~\cite{4HGtri}. The three level system we analyse here differs essentially only in the choice of cavity field which is externally pumped. In these previous two octave systems, this was the field at the lowest frequency. In this work it is the field at the intermediate frequency which is pumped. Just as with the normal OPO and SHG processes, this small change leads to markedly different behaviours.

The system we analyse has the potential to provide enhanced flexibility for quantum interfaces between light and atomic ensembles, quantum state engineering, multiplexing in quantum communications~\cite{multiplex}, the entanglement of atomic ensembles, and quantum teleportation~\cite{Hammerer}. The availability of entanglement and EPR-steering over such a large frequency range will bring further flexibility to the linking of quantum processes at different wavelengths, for example the telecommunications frequencies and atomic systems used in quantum information processing, particularly with regard to quantum memory~\cite{Julsgaard}.

In this article we first provide the Hamiltonian, then develop the equations of motion in the positive-P representation~\cite{P+}. These equations are then solved numerically to find the time evolution of the intracavity fields. We check the full quantum numerical results against those found analytically for the classical steady states, finding that these agree in most parameter regimes. One regime where they do not agree is that in which the classical solutions exhibit self pulsing behaviour. In other regimes we use the steady state solutions for a linearised fluctuation analysis. This allows us to find the oscillation threshold, which is changed from that in the standard OPO. Using the standard input-output relations~\cite{mjc}, we are able to calculate the expressions for squeezing and both bipartite and tripartite EPR steering and inseparability in the output modes. In cases where the output expressions are rather simple, we give these analytically. In other cases the results are produced graphically. We look at the effects of changing the ratio of the two nonlinearities and the cavity damping rates. Finally we examine the effects of an injected signal at the lowest frequency. The range of interesting quantum states found suggests that this system shows promise for emerging quantum technological applications.

\section{Hamiltonian and equations of motion}
\label{sec:Ham}

The system we investigate here uses two $\chi^{(2)}$ nonlinear interactions within the same pumped optical cavity which is resonant for all three frequencies of interest. These could be either two crystals or one customised dielectric~\cite{Zhu} which converts the input field via both up and down conversion. The three interacting electromagnetic fields are the central externally pumped field at frequency $\omega_{2}$, and two others at $\omega_{1}$ and $\omega_{3}$. The field at $\omega_{2}$ interacts via a nonlinearity represented by $\kappa_{1}$ to produce a downconverted field at $\omega_{1}$, where $\omega_{2}=2\omega_{1}$. It also interacts via the nonlinearity represented by $\kappa_{2}$ to produce an upconverted field at $\omega_{3} (=2\omega_{2})$. This field is therefore the fourth harmonic of $\omega_{1}$, with the interacting fields spanning two octaves of frequency difference.

The low frequency field at $\omega _{1}$, is
represented by the bosonic operator $\hat{a}_{1}$. The second harmonic, at $\omega _{2}=2\omega
_{1}$, which will be externally pumped, is represented by $\hat{a}_{2}$, and
the fourth harmonic, at $\omega _{3}=4\omega _{1}$, is represented by $\hat{a}_{3}$. The unitary interaction Hamiltonian in a rotating frame is then
written as 
\begin{equation}
\mathcal{H}_{int}=\frac{i\hbar }{2}\left[ \kappa _{1}(\hat{a}_{1}^{2}\hat{a}%
_{2}^{\dag }-\hat{a}_{1}^{\dag \,2}\hat{a}_{2})+\kappa _{2}(\hat{a}_{2}^{2}%
\hat{a}_{3}^{\dag }-\hat{a}_{2}^{\dag \,2}\hat{a}_{3})\right] .
\label{eq:UHam}
\end{equation}%
Since we are analysing the intracavity configuration, we also have the pumping Hamiltonian, 
\begin{equation}
\mathcal{H}_{pump}=i\hbar \left( \epsilon _{2}\hat{a}_{2}^{\dag }-\epsilon
_{2}^{\ast }\hat{a}_{2}\right) ,  \label{eq:Hpump}
\end{equation}%
where $\epsilon _{2}$ represents an external pumping field which is usually
taken as coherent, although this is not necessary~\cite{Liz}. The damping of
the cavity into a zero temperature Markovian reservoir is described by the Lindblad
superoperator 
\begin{equation}
\mathcal{L}\rho =\sum_{i=1}^{3}\gamma _{i}\left( 2\hat{a}_{i}\rho \hat{a}%
_{i}^{\dag }-\hat{a}_{i}^{\dag }\hat{a}_{i}\rho -\rho \hat{a}_{i}^{\dag }%
\hat{a}_{i}\right) ,  
\label{eq:Lindblad}
\end{equation}%
where $\rho $ is the system density matrix and $\gamma _{i}$ is the cavity
loss rate at $\omega _{i}$. In this work we will treat all three optical
fields as being at resonance with the optical cavity. While including detuning is possible, this makes analytical results very difficult to obtain, so we will stick to the simplest case here. In general, any detuning acts to degrade the correlations used to measure squeezing and entanglement in a $\chi^{(2)}$ system~\cite{Granja}.

In order to analyse this system, we will use the well known and exact quantum phase space method, the positive-P representation~\cite{P+}, which allows us to readily calculate any time-normally-ordered operator moments.
Following the usual procedures~\cite{DFW}, we derive equations of motion
in the positive-P representation~\cite{P+}, 
\begin{eqnarray}
\frac{d\alpha _{1}}{dt} &=&-\gamma _{1}\alpha _{1}+\kappa _{1}\alpha
_{1}^{+}\alpha _{2}+\sqrt{\kappa _{1}\alpha _{2}}\,\eta _{1},  \notag \\
\frac{d\alpha _{1}^{+}}{dt} &=&-\gamma _{1}^{+}\alpha _{1}^{+}+\kappa
_{1}\alpha _{1}\alpha _{2}^{+}+\sqrt{\kappa _{1}\alpha _{2}^{+}}\,\eta _{2},
\notag \\
\frac{d\alpha _{2}}{dt} &=&\epsilon _{2}-\gamma _{2}\alpha _{2}+\kappa
_{2}\alpha _{2}^{+}\alpha _{3}-\frac{\kappa _{1}}{2}\alpha _{1}^{2}+\sqrt{%
\kappa _{2}\alpha _{3}}\,\eta _{3},  \notag \\
\frac{d\alpha _{2}^{+}}{dt} &=&\epsilon _{2}^{\ast }-\gamma _{2}\alpha
_{2}^{+}+\kappa _{2}\alpha _{2}\alpha _{3}^{+}-\frac{\kappa _{1}}{2}\alpha
_{1}^{+\,2}+\sqrt{\kappa _{2}\alpha _{3}^{+}}\,\eta _{4},  \notag \\
\frac{d\alpha _{3}}{dt} &=&-\gamma _{3}\alpha _{3}-\frac{\kappa _{2}}{2}%
\alpha _{2}^{2},  \notag \\
\frac{d\alpha _{3}^{+}}{dt} &=&-\gamma _{3}\alpha _{3}^{+}-\frac{\kappa _{2}}{2}\alpha _{2}^{+\,2}.  
\label{eq:Pplus}
\end{eqnarray}
It should be noted that these have the same form in either It\^{o} or
Stratonovich calculus \cite{SMCrispin}. In the above, the complex variable
pairs $(\alpha _{i},\alpha _{j}^{+})$ correspond to the operator pairs $(\hat{a}_{i},\hat{a}_{j}^{\dag })$ in the sense that stochastic averages of
products converge to normally-ordered operator expectation values, e.g. $\overline{\alpha _{i}^{+\,m}\alpha _{j}^{n}}\rightarrow \langle \hat{a}_{i}^{\dag \,m}\hat{a}_{j}^{n}\rangle $. The $\eta _{j}$ are Gaussian noise
terms with the properties $\overline{\eta _{i}}=0$ and $\overline{\eta
_{j}(t)\eta _{k}(t^{\prime })}=\delta _{jk}\delta (t-t^{\prime })$. Although there can be divergence problems with the positive-P representation, it is known to be accurate where it converges, which is the case with all results presented here.

\section{Steady-state and threshold properties}
\label{sec:cavidade}

In order to obtain analytical steady-state results for the intracavity intensities and amplitudes, we solve the semi-classical equivalents of Eq.~\ref{eq:Pplus}, simply obtained by removing the noise terms. The results thus obtained can be checked against stochastic integration of the full equations. This procedure also allows us to calculate the threshold pumping value at which the downconversion process begins to produce non-zero amplitudes in the low frequency mode. This threshold behaviour is well known from the theory of the optical parametric oscillator (OPO)~\cite{DMW,Arabe}. A stability analysis of the system allows the threshold pumping amplitude to be calculated as 
\begin{equation}
\epsilon _{2}^{c}=\frac{\gamma _{1}\gamma _{2}}{\kappa _{1}}+\frac{\gamma
_{1}^{3}\kappa _{2}^{2}}{2\gamma _{3}\kappa _{1}^{3}}.
\label{eq:critpump}
\end{equation}
We immediately see that this is higher than the threshold for isolated downconversion, where the threshold is 
$\gamma _{1}\gamma _{2}/\kappa _{1}$. The increased pump power is required because the upconversion process to produce the mode at $\omega_{3}$ also depletes the pump in our system.

The steady state amplitudes for the three modes can be found in the two different cases:\\ 
(i) below threshold $\epsilon _{2}<\epsilon _{2}^{c}$,  
\begin{eqnarray}
\alpha _{1}^{ss} &=&0,  \nonumber \\
\alpha _{2}^{ss} &=&\frac{\xi }{3\kappa _{2}^{2}}-\frac{2\gamma _{2}\gamma
_{3}}{\xi },  \nonumber \\
\alpha _{3}^{ss} &=&-\frac{\kappa _{2}\left( \alpha _{2}^{ss}\right) ^{2}}{2\gamma _{3}},
\label{eq:belowcrit}
\end{eqnarray}
where
\begin{equation}
\xi =\left( 27\epsilon _{2}\gamma _{3}\kappa _{2}^{4}+3\sqrt{3}\sqrt{8\gamma _{2}^{3}\gamma _{3}^{3}\kappa _{2}^{6}+27\epsilon _{2}^{2}\gamma
_{3}^{2}\kappa _{2}^{8}}\right) ^{1/3},
\label{eq:xi}
\end{equation}
and \\
(ii) above threshold $\epsilon _{2}>\epsilon _{2}^{c}$,  
\begin{eqnarray}
\alpha _{1}^{ss} &=&\pm \frac{2}{\kappa _{1}}\left( \epsilon
_{2}-\epsilon _{2}^{c}\right)  ,  \nonumber \\
\alpha _{2}^{ss} &=&\frac{\gamma _{1}}{\kappa _{1}}, \nonumber \\
\alpha _{3}^{ss} &=& -\frac{\gamma _{1}^{2}\kappa _{2}}{2\kappa _{1}^{2}\gamma
_{3}}.
\label{eq:abovecrit}
\end{eqnarray}%
As with the standard OPO, the system exhibits similar behaviour to a second-order phase transition at 
$\epsilon _{2}=\epsilon _{2}^{c}$. When the pumping is above threshold, the 
below-threshold solution for the fundamental frequency field $\alpha _{1}^{ss}=0$
becomes unstable and the system moves onto a new stable
branch witht two solutions of the the fundamental field
having equal amplitude and opposite phase.
The steady amplitudes of the central frequency $\omega _{2}$ and higher frequency $\omega _{3}$ 
modes have opposite phases whether the system is running below or above threshold. What is noticeable is that the steady state solutions above threshold for $\alpha_{2}$ and $\alpha_{3}$ have no dependence on the pump power. Once the cavity is being pumped above the oscillation threshold, these two fields do not change with changes in the pumping.

\begin{figure}[htbp]
\begin{center}
\includegraphics[width=0.85\columnwidth]{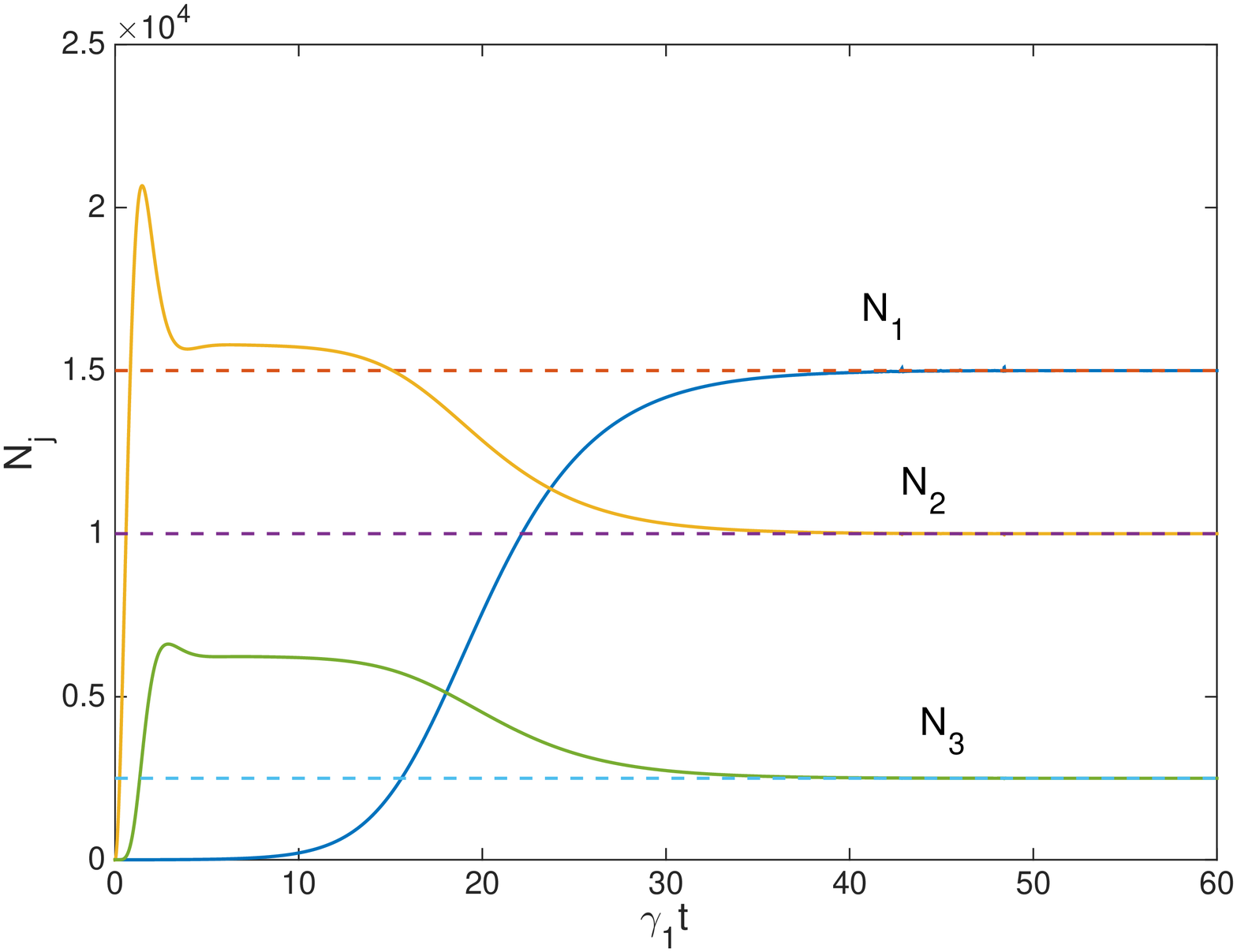}
\end{center}
\caption{(colour online) The intracavity intensities calculated via $4\times 10^{5}$ trajectories of the positive-P equations are shown as the solid lines. The dashed lines are the analytical steady-state expressions. The parameters used are $\gamma_{j}=1$, $\kappa_{1}=\kappa_{2}=10^{-2}$, and $\epsilon=1.5\epsilon_{c}$. Averaging errors are smaller than the plotted linewidths. All quantities plotted in this and subsequent graphics are dimensionless.}
\label{fig:intensities}
\end{figure}

The time development of the intensities above threshold is shown in Fig.~\ref{fig:intensities} in the fully 
quantum picture with the positive-P equations integrated over $4\times 10^{5}$ stochastic trajectories. 
With $\epsilon _{2}=1.5\epsilon _{2}^{c}$ we see that the analytical steady-state values, plotted as dashed lines, are in good agreement with the quantum solutions.

\begin{figure}[htbp]
\begin{center}
\includegraphics[width=0.85\columnwidth]{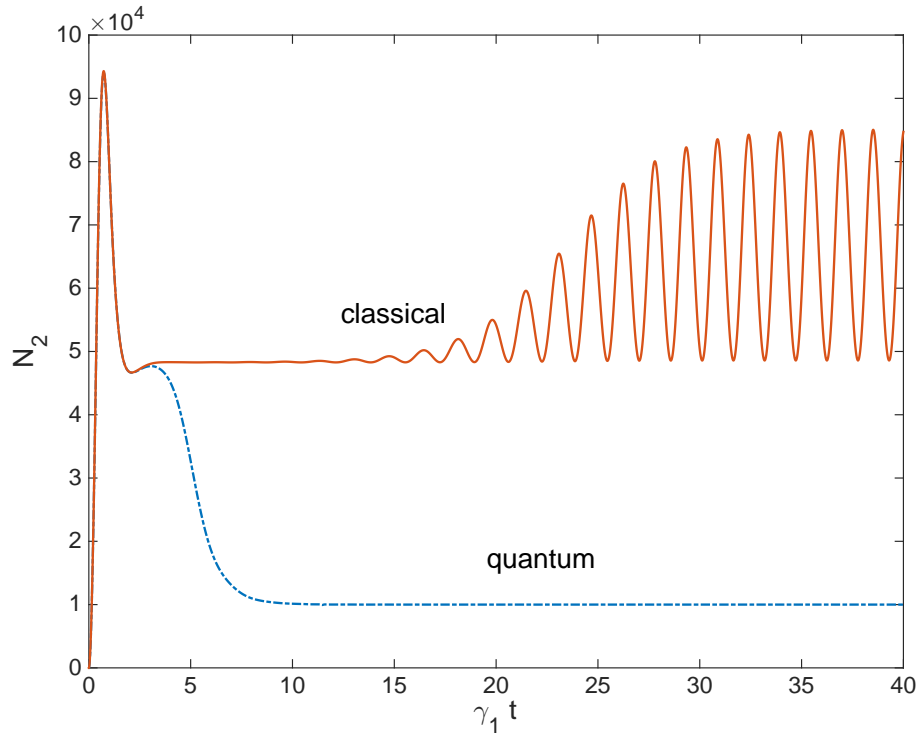}
\end{center}
\caption{(colour online) The classical and quantum solutions for $N_{2}$, with the same parameters as Fig.~\ref{fig:intensities} except for $\epsilon_{2}=5\epsilon_{2}^{c}$. Both integrations have a small complex seed in the initial conditions, with $\alpha_{1}(0)=0$, $\alpha_{2}(0)=1+2i$ and $\alpha_{3}(0)=1-2i$.}
\label{fig:selfpulse}
\end{figure}

It is also well known that in normal second harmonic generation (SHG) there is a pumping threshold above which the output intensities exhibit a periodic pulsing behaviour~\cite{pulse,Bache}. In the present case the classical behaviour of the system is similar and a hard mode transition can be found above which self-pulsing occurs. However, this does not survive the full quantum treatment, with the oscillations disappearing completely. A less pronounced damping of self-pulsing oscillations has recently been found in a full quantum treatment of other cascaded systems~\cite{3HG,4HG} and shows the dangers of relying on classical analyses of quantum optical systems. The canonical method to calculate self pulsing in SHG is to numerically integrate the classical equations with a small complex seed in one or both the modes. Without this seed, the self-pulsing is not found, although it appears with integration of the positive-P equations without needing any seed at all.  For our system, small complex seeds in the initial condition of the classical simulations gives self-pulsing, as shown in Fig.~\ref{fig:selfpulse}. On the other hand, the quantum solution diverges from this at short times, to enter a steady state with a much lower average value. The reason for this is that the classical solutions stay on the unstable branch of the solutions for $\alpha_{1}$, remaining at zero. The classical solution is unphysical. In the quantum case, spontaneous downconversion early in the evolution leads to stimulated downconversion and the steady state remains on the stable branch. A small injected signal $\epsilon_{1}$ in the classical integration will also push the solutions onto the stable branch, and in this case self-pulsing is found neither classically nor quantum mechanically.  

\section{Ornstein-Uhlenbeck analysis and fluctuation spectra}
\label{sec:OU}

When nonlinear optical media are held inside a pumped optical cavity, the
accessible observables are usually the output spectral correlations, which
are accessible using homodyne measurement techniques~\cite{mjc}. These are
readily calculated in the steady state by treating the system as an
Ornstein-Uhlenbeck process~\cite{SMCrispin}. In order to do this, we begin
by expanding the positive-P variables into their steady-state expectation
values plus delta-correlated Gaussian fluctuation terms, e.g. 
\begin{equation}
\alpha_{ss} \rightarrow \langle\hat{a}\rangle_{ss}+\delta\alpha.
\label{eq:fluctuate}
\end{equation}
Given that we can calculate the $\langle\hat{a}\rangle_{ss}$, we may now
write the equations of motion for the fluctuation terms. The resulting
equations are written for the vector of fluctuation terms as 
\begin{equation}
d\delta \vec{\alpha}=-A\delta \vec{\alpha}dt+Bd\vec{W},
\label{eq:OEeqn}
\end{equation}
where $A$ is the drift matrix containing the steady-state solution, $B$ is
found from the factorisation of the drift matrix of the original
Fokker-Planck equation, $D=BB^{T}$, with the steady-state values substituted
in, and $d\vec{W}$ is a vector of Wiener increments. As long as the matrix $A
$ has no eigenvalues with negative real parts, this method may be used to
calculate the intracavity spectra via 
\begin{equation}
S(\omega) = (A+i\omega)^{-1}D(A^{\mbox{\small{T}}}-i\omega)^{-1},
\label{eq:Sout}
\end{equation}
from which the output spectra are calculated using the standard input-output
relations~\cite{mjc}.

In this case, $A$ is found as
\begin{equation}
A = 
\begin{bmatrix}
\gamma_{1} & -\kappa_{1}\alpha_{2} & -\kappa_{1}\alpha_{1}^{\ast} & 0 & 0 & 0
\\ 
-\kappa_{1}\alpha_{2}^{\ast} & \gamma_{1} & 0 & -\kappa_{1}\alpha_{1} & 0 & 0
\\ 
\kappa_{1}\alpha_{1} & 0 & \gamma_{2} & -\kappa_{2}\alpha_{3} & 
-\kappa_{2}\alpha_{2}^{\ast} & 0 \\ 
0 & \kappa_{1}\alpha_{1}^{\ast} & -\kappa_{2}\alpha_{3}^{\ast} & \gamma_{2}
& 0 & -\kappa_{2}\alpha_{2} \\ 
0 & 0 & \kappa_{2}\alpha_{2} & 0 & \gamma_{3} & 0 \\ 
0 & 0 & 0 & \kappa_{2}\alpha_{2}^{\ast} & 0 & \gamma_{3}
\end{bmatrix}
\label{eq:Amat}
\end{equation}
and $D$ is a $6\times 6$ matrix with 
$\left[\kappa_{1}\alpha_{2},\kappa_{1}\alpha_{2}^{\ast},\kappa_{2}\alpha_{3},\kappa_{2}\alpha_{3}^{\ast},0,0\right]$ 
on the diagonal. In the above, the $\alpha_{j}$ should be read as their
steady-state mean values, so that $\alpha _{j}^{\ast }=\overline{\alpha _{j}^{+}}$, for example. 
These are now complex numbers that are the 
averages of the positive-P stochastic variables. Because we have parametrised
our system using $\gamma_{1}=1$, the frequency $\omega$ is in units of $\gamma_{1}$. 
$S(\omega)$ is now in terms of quadratic products of the fluctuation operators such as
$\delta\alpha_{i}\delta\alpha_{j}$ and $\delta\alpha_{i}^{\ast}\delta\alpha_{j}^{\ast}$.

Since quadrature properties are what is measured by homodyne detection, we define the amplitude and phase quadrature operators as
\begin{equation}
\begin{array}[b]{l}
\hat{X}_{j}=\hat{a}_{j}+\hat{a}_{j}^{\dag }, \\
\hat{Y}_{j}=-i\left( \hat{a}_{j}-\hat{a}_{j}^{\dag }\right) .
\end{array}
\end{equation}
We note here that other definitions are sometimes used in the literature and that this changes the numerical value of the Heisenberg uncertainty principle. Our choice gives $V(\hat{X}_{j})V(\hat{Y}_{j}\geq 1$ and means that squeezing in a particular quadrature exists whenever its variance is found to be less than 1.

To express the fluctuation expressions in terms of the canonical quadratures, we calculate 
\begin{equation}
S^{q}\left( \omega \right) =QSQ^{T},
\label{eq:quadtransform}
\end{equation}
where $Q$ is the block diagonal $6\times 6$ matrix constructed from 
\begin{equation}
q=\left[
\begin{array}{cc}
1 & 1 \\
-i & i%
\end{array}%
\right].
\end{equation}%
$S^{q}\left( \omega \right)$ gives us the products from which we construct the
output variances and covariances for modes $i$ and $j$ as, 
\begin{eqnarray}
V\left( \hat{X}_{i},\hat{X}_{j}\right) &=& \delta _{ij}+\sqrt{\gamma _{i}\gamma
_{j}}\left( S_{2i-1,2j-1}^{q}+S_{2j-1,2i-1}^{q}\right), \nonumber \\
V\left( \hat{Y}_{i},\hat{Y}_{j}\right) &=& \delta _{ij}+\sqrt{\gamma _{i}\gamma
_{j}}\left( S_{2i,2j}^{q}+S_{2j,2i}^{q}\right),
\label{eq:quadvars}
\end{eqnarray}
in which the variances and covariances are defined as 
$V\left( \hat{X}_{i}\right) =\left\langle \hat{X}_{i}^{2}\right\rangle -\left\langle
\hat{X}_{i}\right\rangle^{2}$ and $V\left( \hat{X}_{i},\hat{X}_{j}\right) =\left\langle \hat{X}_{i}\hat{X}_{j}\right\rangle -\left\langle \hat{X}_{i}\right\rangle \left\langle \hat{X}
_{j}\right\rangle.$

\section{Steady state bipartite correlations}
\label{sec:sscorrelations}

\begin{figure}[htbp]
\begin{center}
\includegraphics[width=0.85\columnwidth]{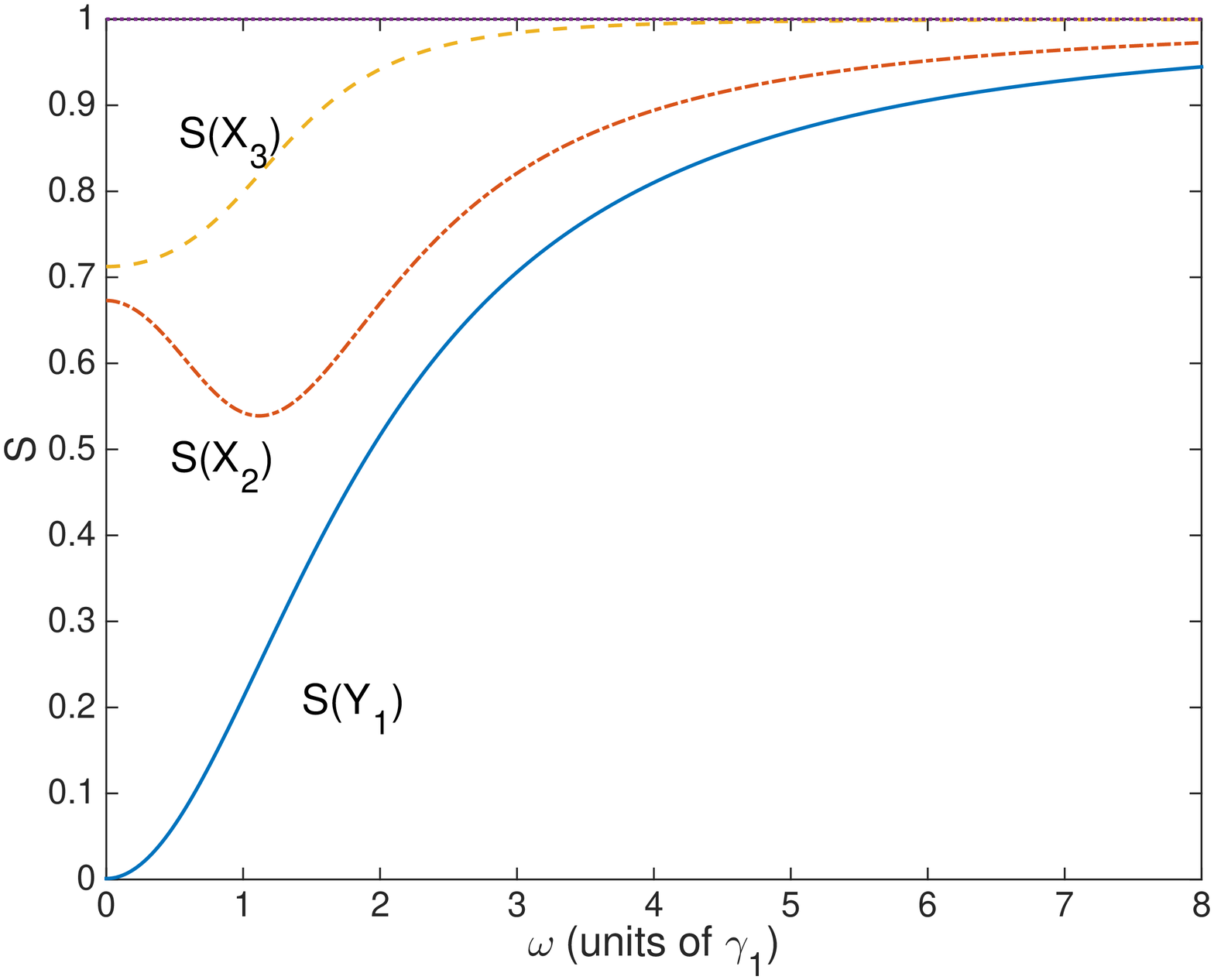}
\end{center}
\caption{(colour online) Quadrature variances for the three squeezed quadratures below threshold, with 
$\kappa _{1}=\kappa _{2}=0.01$, $\gamma _{1}=\gamma _{2}=\gamma _{3}=1$, and 
$\epsilon _{2}=0.9\epsilon _{2}^{c}$. The frequency axis is in units of the linewidth of the 
fundamental, $\gamma _{1}$.}
\label{fig:squeeze}
\end{figure}

The squeezing in the amplitude and phase 
quadrature for the three different modes can be calculated analytically following
from Eq.~\ref{eq:belowcrit} and Eq.~\ref{eq:quadtransform}. 
Since the fundamental mode has a mean amplitude of zero
below threshold, this simplifies the drift matrix and we can 
derive the below threshold output squeezing spectra as 
\begin{eqnarray}
S_{1\pm }\left( \omega \right)  &=&1\pm \frac{4\gamma _{1}\kappa
_{1}\alpha _{2}}{\omega ^{2}+\left( \gamma _{1}\mp \kappa _{1}\alpha
_{2}\right) ^{2}},  \nonumber \\
S_{2\pm }\left( \omega \right)  &=&1\pm 4\gamma _{2}\kappa _{2}\alpha
_{3}\left( \omega ^{2}+\gamma _{3}^{2}\right) \eta _{\pm }\left( \omega
\right),   \nonumber \\
S_{3\pm }\left( \omega \right)  &=&1\pm 4\gamma _{3}\alpha
_{2}^{2}\alpha _{3}\kappa _{2}^{3}\eta _{\pm }\left( \omega \right),
\label{eq:quadspekbelow}
\end{eqnarray}
where 
\begin{equation}
\eta _{\pm }\left( \omega \right) =\frac{1}{\omega^{2}\left( \gamma
_{2}+\gamma _{3}\mp \kappa _{2}\alpha _{3}\right)^{2}+\left( -\omega
^{2}+\kappa _{2}^{2}\alpha _{2}^{2}+\gamma _{2}\gamma _{3}\mp \gamma
_{3}\kappa _{2}\alpha _{3}\right)^{2}},
\label{eq:eta}
\end{equation}
and
$S_{j+}\left(\omega \right) =S\left(X_{j}\right) ,S_{j-}\left(\omega
\right)=S\left(Y_{j}\right).$  The spectral variances of the squeezed quadratures are shown in Fig.~\ref{fig:squeeze}, for $\epsilon_{2}=0.9\epsilon_{2}^{c}$. We note here that all spectra shown are symmetric about zero frequency. What we notice is that the quadratures which exhibit squeezing are those we expect from parametric downconversion, with $\hat{Y}_{1}$ being squeezed, and from second harmonic generation, with both $\hat{X}_{2}$ and $\hat{X}_{3}$ being squeezed.

\begin{figure}[htbp]
\begin{center}
\includegraphics[width=0.85\columnwidth]{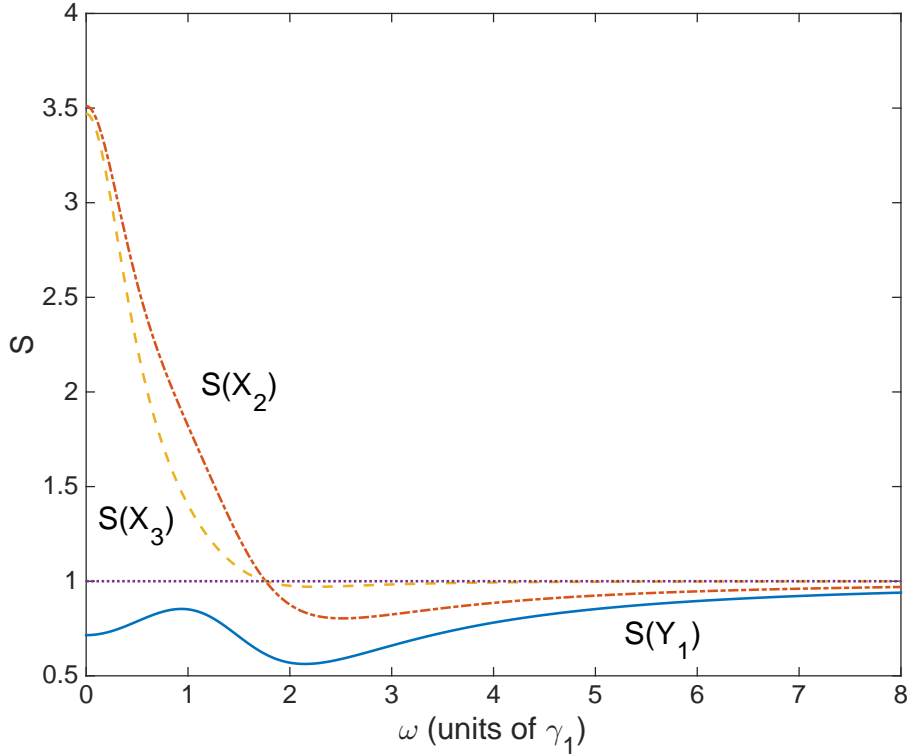}
\end{center}
\caption{(colour online) Quadrature variances for the three squeezed quadratures above threshold, with 
$\kappa _{1}=\kappa _{2}=0.01$, $\gamma _{1}=\gamma _{2}=\gamma _{3}=1$, and 
$\epsilon _{2}=1.5\epsilon _{2}^{c}$. The dotted line at one is a guide to the eye and the frequency axis is in units of the linewidth of the 
fundamental, $\gamma _{1}$.}
\label{fig:squeezeabove}
\end{figure}

Above threshold the analytical expressions for the output squeezing are quite lengthy, mainly due to that fact that the low frequency mode now has a non-zero solution. We will not give these here, but will illustrate the results in Fig.~\ref{fig:squeezeabove}, for $\epsilon_{2}=1.5\epsilon_{2}^{c}$. We see that the same quadratures are squeezed as below threshold, but that the degree of squeezing has been reduced.

The next question we raise is whether any of the possible bipartitions will exhibit the Einstein-Podolsky-Rosen paradox~\cite{EPR}, now commonly known as EPR steering~\cite{Erwin,Jonesteer}. In the continuous variable case, this is usually measured using the Reid inequalities for the inferred variances~\cite{EPRMDR,ZYOu}. This is written for the output spectral variances as 
\begin{equation}
EPR_{ij}(\omega) = S^{inf}(\hat{X}_{i})S^{inf}(\hat{Y}_{i})\geq 1,
\label{eq:eprMDR}
\end{equation}
where 
\begin{eqnarray}
S_{inf}(\hat{X}_{i}) &=& S(\hat{X}_{i})-\frac{[S(\hat{X}_{i},\hat{X}_{j})]^{2}}{S(\hat{X}_{j})},  \nonumber \\
S_{inf}(\hat{Y}_{i}) &=& S(\hat{Y}_{i})-\frac{[S(\hat{Y}_{i},\hat{Y}_{j})]^{2}}{S(\hat{Y}_{j})}. 
\label{eq:EPRdef}
\end{eqnarray}

In the language of EPR-steering, $EPR_{ij}<1$ shows that mode $i$ can be steered by measurements of mode $j$. In some cases asymmetric steering is possible, where $EPR_{ij}<1$ while $EPR_{ji}>1$. The question as to whether this was possible was first raised by Wiseman \etals~\cite{Wiseman}, and answered in the affirmative for Gaussian measurements by Olsen and Bradley~\cite{SFG}, Midgley \etals~\cite{sapatona}, and H\"andchen \etals~\cite{Handchen}. It has since been shown that asymmetric steering is generally possible~\cite{Bowles}, without any restriction on measurements. 
Because EPR steerable states are a strict subset of the entangled states, 
both symmetric and asymmetric steering demonstrate that the two modes concerned 
are fully bipartite entangled. We will therefore use the Reid inequalities to demonstrate both EPR steering and bipartite entanglement. 

We obtain the below threshold covariances between each pair of modes as
\begin{eqnarray}
S(\hat{X}_{1},\hat{X}_{2}) &=& S(\hat{X}_{1},\hat{X}_{3})=0,  \nonumber \\
S(\hat{Y}_{1},\hat{Y}_{2}) &=& S(\hat{Y}_{1},\hat{Y}_{3})=0,  \nonumber \\
S(\hat{X}_{2},\hat{X}_{3}) &=& -4\alpha _{2}\alpha _{3}\gamma _{3}\sqrt{\gamma _{2}\gamma _{3}}\kappa _{2}^{2}\eta _{+}\left( \omega \right) ,
\nonumber \\
S(\hat{Y}_{2},\hat{Y}_{3}) &=& 4\alpha _{2}\alpha _{3}\gamma _{3}\sqrt{\gamma
_{2}\gamma _{3}}\kappa _{2}^{2}\eta _{-}\left( \omega \right).
\label{eq:covariances}
\end{eqnarray}
Since the covariances between modes 1 and 2 and modes 1 and 3 are zero, we
can easily find four of the possible EPR  correlations as
\begin{eqnarray}
EPR_{12} &=& EPR_{13} = S_{1+}\left( \omega \right)S_{1-}\left( \omega \right), \nonumber \\
EPR_{21} &=& S_{2+}\left( \omega \right) S_{2-}\left( \omega \right), \nonumber \\
EPR_{31} &=& S_{3+}\left( \omega \right) S_{3-}\left( \omega \right).
\label{eq:EPRanalytic}
\end{eqnarray}
It is obvious that none of these bipartitions can exhibit EPR steering below threshold, due to to Heisenberg Uncertainty Principal. An interesting result is that, although $EPR_{21}$ and $EPR_{31}$ are products of variances for different modes, they have equal values, with neither falling below one. This is not the case above threshold, where these two are no longer equal.

\begin{figure}[htbp]
\begin{center}
\includegraphics[width=0.85\columnwidth]{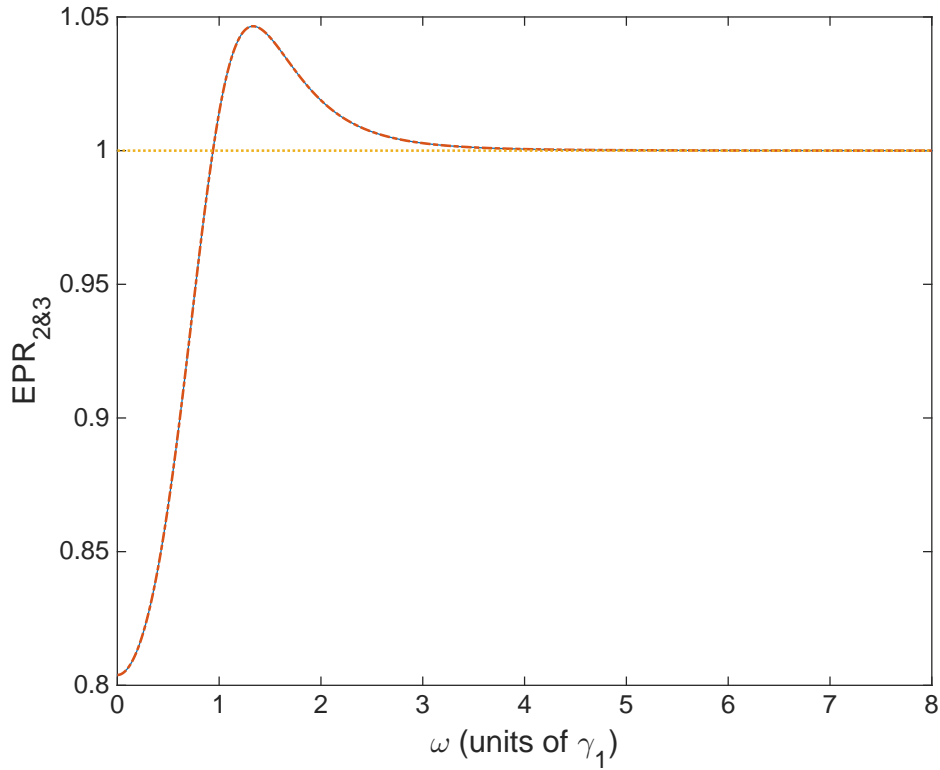}
\end{center}
\caption{(colour online) $EPR_{23}$ and $EPR_{32}$ for 
$\kappa _{1}=\kappa _{2}=0.01$, $\gamma _{1}=\gamma _{2}=\gamma _{3}=1$, and 
$\epsilon _{2}=0.9\epsilon _{2}^{c}$. The dotted line at one is a guide to the eye.}
\label{fig:EPR23below}
\end{figure}

The case for modes 2 and 3, however, is different. A complicated analytical expression tells us that
$EPR_{32}=EPR_{23}$, so that any EPR steering here is completely symmetric. The result for the same parameters as in Fig.~\ref{fig:squeeze} is shown in Fig.~\ref{fig:EPR23below}. We see that the Reid inequalities are violated over a range near zero frequency, meaning that modes 2 and 3 are genuinely bipartite entangled. 

Above threshold, the analytical expressions for all bipartitions become extremely complicated, and are best represented graphically.
We will begin with $\kappa_{1}=\kappa_{2}$ and all cavity loss rates being equal, showing the effects of varying these later in the article.
We find that modes 1 and 2 exhibit symmetric EPR steering over a broad range, while 1 and 3 exhibit completely asymmetric EPR steering over a narrower range of frequencies. The two higher frequency modes, which exhibit EPR steering below threshold, lose this property completely as the solution for $\alpha_{1}$ moves onto the stable branch where it has non-zero amplitude. In terms of entanglement and EPR steering properties, the system changes completely at threshold.

\begin{figure}[htbp]
\begin{center}
\includegraphics[width=0.85\columnwidth]{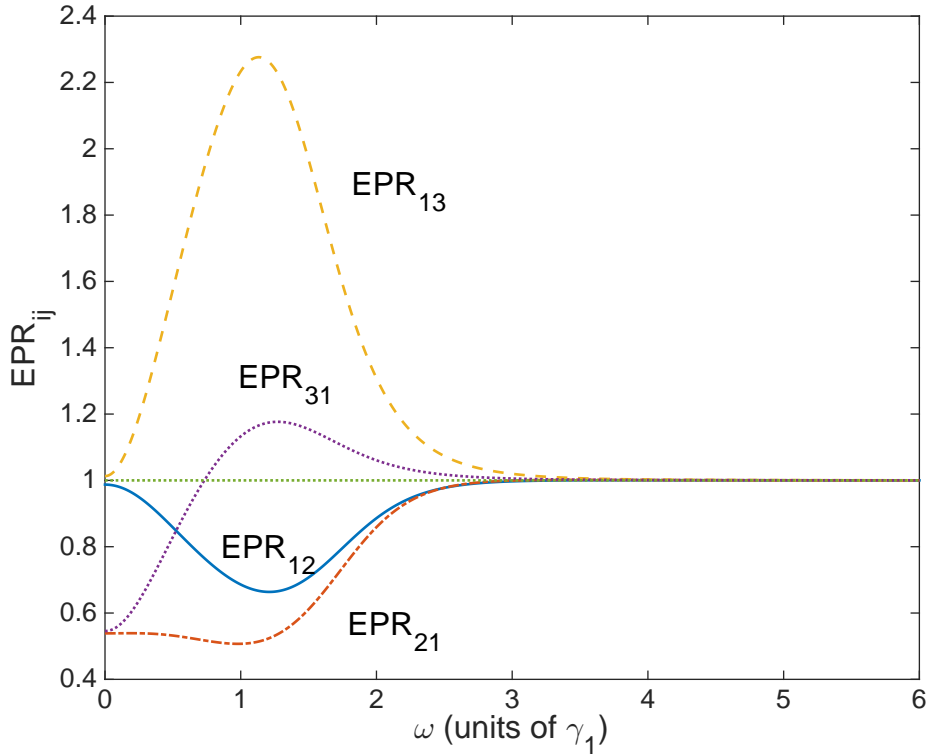}
\end{center}
\caption{(colour online) The EPR correlations which violate the inequality above threshold, for 
$\kappa _{1}=\kappa _{2}=0.01$, $\gamma _{1}=\gamma _{2}=\gamma _{3}=1$, and 
$\epsilon _{2}=1.5\epsilon _{2}^{c}$. The dotted line at one is a guide to the eye.}
\label{fig:EPRabove}
\end{figure}

\begin{figure}[htbp]
\begin{center}
\includegraphics[width=0.85\columnwidth]{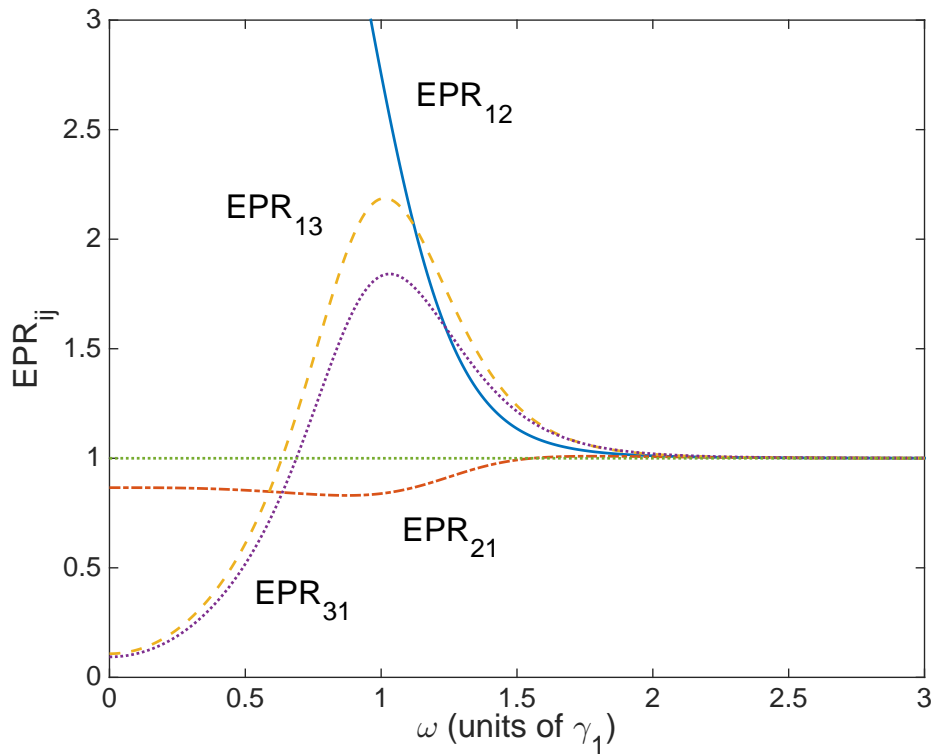}
\end{center}
\caption{(colour online) The EPR correlations which violate the inequality above threshold, for 
$\kappa _{1}=\kappa _{2}=0.01$, $\gamma _{1}=\gamma _{3} = 1 = 10\gamma _{2}$, and 
$\epsilon _{2}=1.5\epsilon _{2}^{c}$. The dotted line at one is a guide to the eye.}
\label{fig:EPRsmallg2}
\end{figure}

We find that the symmetry or asymmetry of the EPR steering between the output modes above threshold can be 
simply controlled by the ratio of loss rates and the  ratio of nonlinearities.
Firstly, in Fig.~\ref{fig:EPRsmallg2}, we show the results of a loss rate for the middle frequency which is one tenth of that for the other two, i.e. $\gamma_{2}=0.1\gamma_{1}=0.1\gamma_{3}$. Whereas modes 1 and 2 exhibited symmetric steering for equal loss rates, their steering is now asymmetric. The opposite has happened with modes 1 and 3, with their steering now being symmetric.
The symmetry properties of the EPR steering can be controlled by adjusting the cavity loss rates, as was also found with intracavity second harmonic generation~\cite{SHGEPR}.   
 
\begin{figure}[htbp]
\begin{center}
\includegraphics[width=0.85\columnwidth]{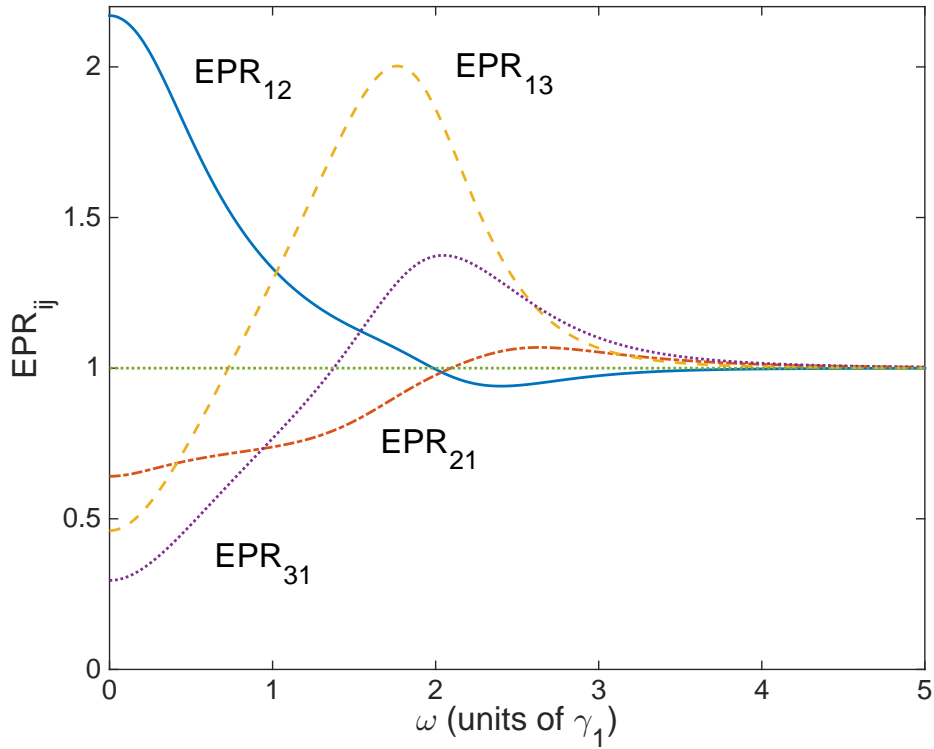}
\end{center}
\caption{(colour online) The EPR correlations which violate the inequality above threshold, for 
$\gamma _{j} = 1 \forall j$,
$\epsilon _{2}=1.5\epsilon _{2}^{c}$ and $\kappa_{2}=1.5\kappa_{1}$, with $\kappa_{1}=0.01$. The dotted line at one is a guide to the eye.}
\label{fig:EPRkappa}
\end{figure}

Changing the ratio $\kappa_{1}/\kappa_{2}$ also has an effect on the EPR steering properties above threshold. We can see in Fig.~\ref{fig:EPRkappa}  that this can result in asymmetric steering in the bipartition of modes 1 and 2, with this swapping over at a certain frequency. Below $\omega \approx 2.1\gamma_{1}$, mode 1 can steer mode 2, while above this frequency there is a small violation of the inequality by $EPR_{12}$. The pairing of 1 and 3 exhibits both symmetric and asymmetric EPR steering as the measurement frequency changes.
We did not find any any steering involving the pair of fields at $\omega_{2}$ and  $\omega_{3}$,
for the whole parameter range investigated with this ratio of the nonlinearities.

\section{Tripartite correlations}
\label{sec:tri}

\begin{figure}[htbp]
\begin{center}
\includegraphics[width=0.85\columnwidth]{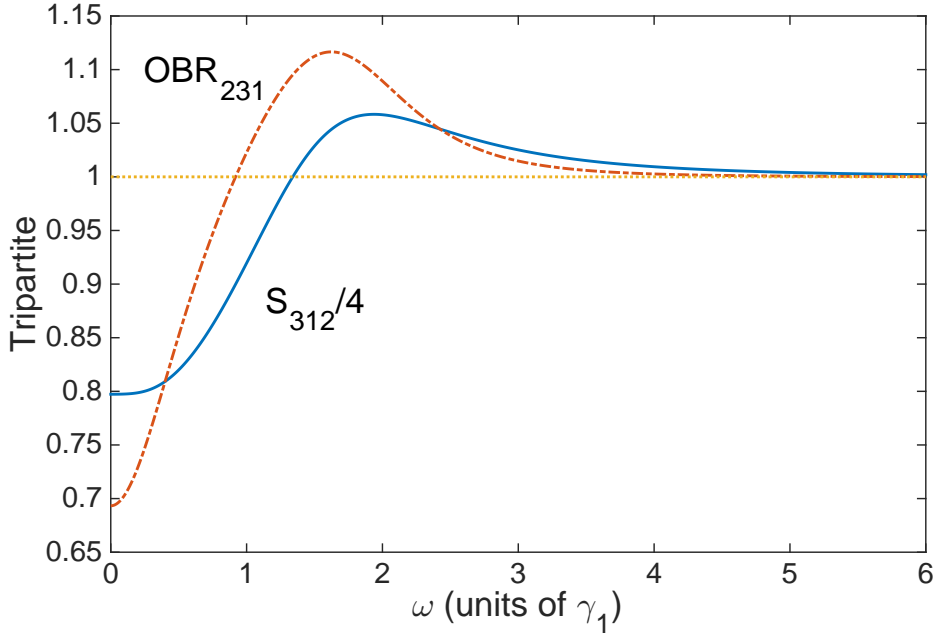}
\end{center}
\caption{(colour online) The spectral tripartite correlations which violate the inequalities above threshold, for the parameters
$\gamma _{j} = 1 \forall j$,
$\epsilon _{2}=1.5\epsilon _{2}^{c}$ and $\kappa_{2}=\kappa_{1}=0.01$. The dotted line at one is a guide to the eye.}
\label{fig:tripart}
\end{figure}

There are several methods of detecting tripartite inseparability and entanglement, with one common technique being based on inequalities developed by van Loock and Furusawa (vLF)~\cite{vLF}. These have proven useful for other cascaded 
systems~\cite{AxMuzzJPB,AxMuzz}. The spectral inequalities we will use here are the set 
\begin{equation}
S_{ijk} = S(\hat{X}_{i}-\frac{\hat{X}_{j}+\hat{X}_{k}}{\sqrt{2}})+S(\hat{Y}_{i}+\frac{\hat{Y}_{j}+\hat{Y}_{k}}{\sqrt{2}}) \geq 4,
\label{eq:VLFijk}
\end{equation}
the violation of any one of which is sufficient to prove bipartite inseparability.
Following the work of Teh and Reid~\cite{Teh&Reid}, any one of these less than $2$ demonstrates genuine tripartite entanglement, while one of them less than $1$ demonstrates genuine tripartite EPR steering. We did not find a violation of these inequalities below threshold. Above threshold we found that some, but not all, of the set of inequalities are violated for particular parameter regimes, as shown in Fig.~\ref{fig:tripart}, where we have divided the values of $S_{312}$ by four so as to be directly comparable with the tripartite EPR steering inequality to be described below. This value of $S_{312}$ demonstrates tripartite inseparability for the system.

With our three mode system, investigating tripartite EPR-steering is also of interest. It has been shown by Wang \etals~\cite{Wang} that, in a multipartite system,
the steering of a given quantum mode is allowed when not less than half of the total number
of modes take part in the steering group. In a tripartite system, this means that measurements on
two of the modes are needed to steer the third. 
In order to quantify this, we will use the correlation 
functions developed by Olsen, Bradley, and Reid~\cite{OBR}. 
With spectral tripartite inferred variances defined as
\begin{eqnarray}
S_{inf }^{(t)}\left( \hat{X}_{i}\right)  &=& S\left( \hat{X}_{i}\right) -\frac{\left [ S\left( \hat{X}_{i},\hat{X}_{j}\pm \hat{X}_{k}\right)
\right]^{2} }{S\left( \hat{X}_{j}\pm \hat{X}_{k}\right) },  \nonumber \\
S_{inf }^{(t)}\left( \hat{Y}_{i}\right)  &=& S\left( \hat{Y}_{i}\right) -\frac{\left [ S\left( \hat{Y}_{i},\hat{Y}_{j}\pm \hat{Y}_{k}\right)
\right ]^{2} }{S\left( \hat{Y}_{j}\pm \hat{Y}_{k}\right) },
\label{eq:OBRinf}
\end{eqnarray}
we define
\begin{equation}
OBR_{ijk}=S_{inf }^{(t)}\left( \hat{X}_{i}\right) S_{inf }^{(t)}\left(
\hat{Y}_{i}\right),
\label{eq:OBRproduct}
\end{equation}
so that a value of less than one means that there is an inferred violation  of the Heisenberg uncertainty principal and mode $i$ can be steered by the combined forces of modes $j$ and $k$. 
According to the work of He and Reid~\cite{HeReid}, genuine tripartite steering is demonstrated whenever
\begin{equation}
OBR_{ijk}+OBR_{jki}+OBR_{kij} < 1.
\label{eq:genuinetristeer}
\end{equation}
We did not find genuine tripartite steering for this system. As shown in Fig.~\ref{fig:tripart}, we found that modes 1 and 2 could combine for some parameters to steer mode 3. We investigated a wide parameter regime numerically, but did not find any for which more than one of the modes could be steered by the remaining pair simultaneously.

\section{An injected signal at the lower frequency}
\label{sec:inject}

It is also possible to pump one of the cavity modes other than that at $\omega_{2}$. The process of optical parametric downconversion with an injected signal has been experimentally and theoretically studied in some depth~\cite{Bjorkholm,Haub,Hovde,Plusquellic}, with the injected signal often used for frequency stabilisation. An injected signal has also been shown to have a strong effect on any quantum correlations~\cite{kaled}, both changing the quadratures where squeezing is found and allowing for control of the asymmetry of EPR steering~\cite{signal}. For these reasons, we will examine here the effects of injecting a coherent signal at $\omega_{1}$. Theoretically, this involves another term in the pumping Hamiltonian, so that 
\begin{equation}
{\cal H}_{pump}^{(s)} = {\cal H}_{pump}+i\hbar \left( \epsilon _{1}\hat{a}_{1}^{\dag }-\epsilon
_{1}^{\ast }\hat{a}_{1}\right),
\label{eq:Hpumpsignal}
\end{equation} 
where ${\cal H}_{pump}^{(s)}$ is the pumping Hamiltonian with injected signal. This change means that the equations of motion for $\alpha_{1}$ and $\alpha_{1}^{+}$ will have $\epsilon_{1}$ and $\epsilon_{1}^{\ast}$ added to them.

\begin{figure}[htbp]
\begin{center}
\includegraphics[width=0.85\columnwidth]{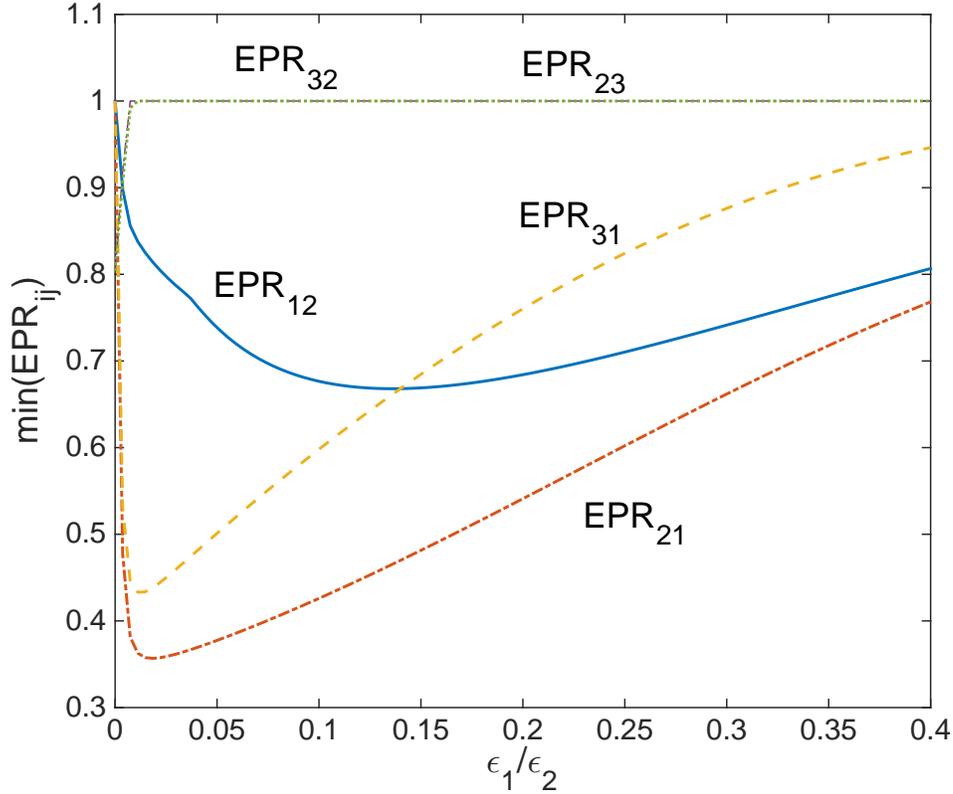}
\end{center}
\caption{(colour online) The minima of the spectral bipartite EPR steering correlations with injected signal which violate the inequality, for parameters
$\gamma _{j} = 1 \forall j$,
$\epsilon _{2}=0.9\epsilon _{2}^{c}$, and $\kappa_{2}=\kappa_{1}=0.01$. The dotted line at one is a guide to the eye.}
\label{fig:finject}
\end{figure}

The immediate effect of an injected signal is to change the threshold properties of the system, with the low frequency mode developing a steady-state non-zero amplitude for all finite values of $\epsilon_{1}$. There is now no critical pump value for $\epsilon_{2}$, with the solutions remaining on the stable branch for all pumping values. The injected signal has an even more dramatic effect on the EPR steering properties of the system. As seen above in Fig.~\ref{fig:EPR23below} the only two modes exhibiting EPR steering below threshold without injected signal were modes 2 and 3. With injected signal, the EPR steering of this bipartition soon vanishes as the signal is increased, which can be seen on the left hand side of Fig.~\ref{fig:finject}, which shows the $EPR_{ij}$ results for steerable bipartitions as the amplitude of the injected signal is increased. The quantities plotted  are the minimum values of the Reid EPR correlations across all frequencies, ($0\leq \omega \leq 6$ numerically), so that a value of one means that the values near the carrier frequency can actually be larger than one. The addition of even a small injected signal (by comparison with $\epsilon_{2}$) has a dramatic effect on the $(1,2)$ and $(1,3)$ bipartitions, These become highly steerable for small injection and then less so as $\epsilon_{1}$ is increased. While $(1,2)$ exhibits symmetric steering, $(1,3)$ is totally asymmetric for these parameters, with $EPR_{31}\leq 1 \leq EPR_{13}$ across the whole range shown. The steerability of $(2,3)$ disappears on the same sort of scale of injection with which the others increase.  

\begin{figure}[htbp]
\begin{center}
\includegraphics[width=0.85\columnwidth]{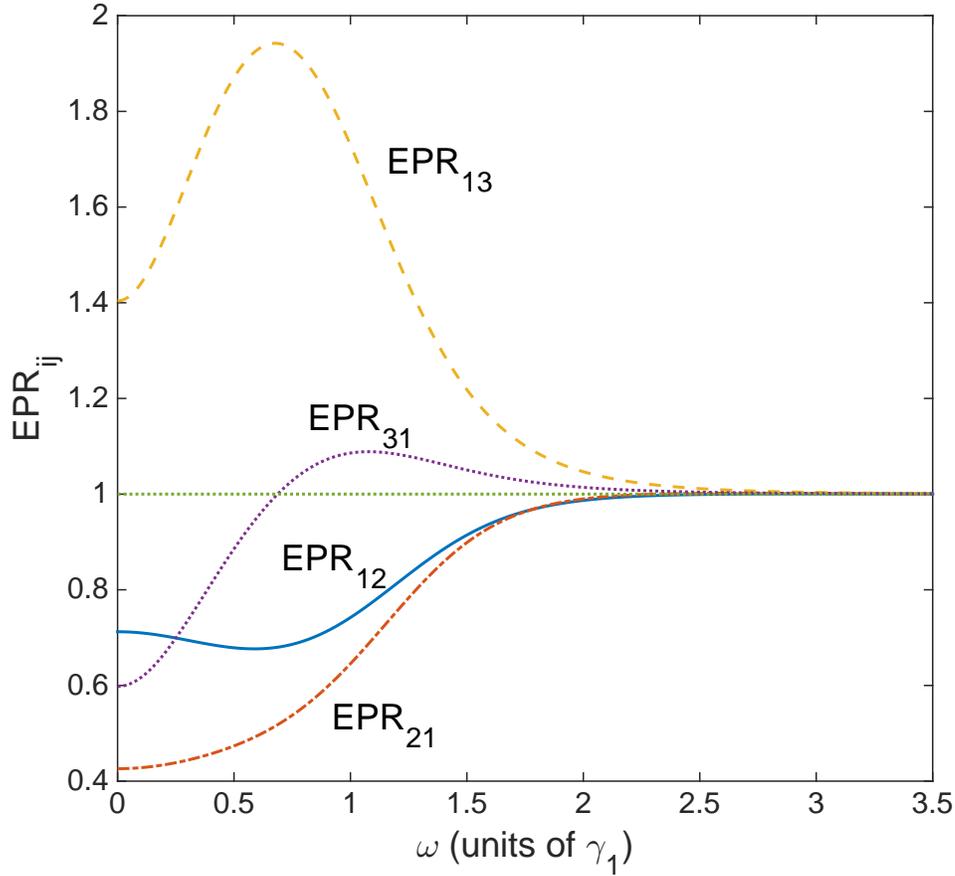}
\end{center}
\caption{(colour online) The spectral bipartite EPR steering correlations with injected signal, for parameters
$\gamma _{j} = 1 \forall j$,
$\epsilon _{2}=0.9\epsilon _{2}^{c}$, $\epsilon_{1}=0.1\epsilon_{2}$, and $\kappa_{2}=\kappa_{1}=0.01$. The dotted line at one is a guide to the eye.}
\label{fig:EPRinject}
\end{figure}

The spectral values of the Reid EPR correlations for the bipartitions which exhibit steering for similar parameters as in 
Fig.~\ref{fig:finject}, but at a fixed $\epsilon_{1}=0.1\epsilon_{2}$, are shown in Fig.~\ref{fig:EPRinject}. The asymmetry of the EPR steering demonstrated by $EPR_{31}$ and $EPR_{13}$ is clearly shown. Nevertheless, this result shows that modes 1 and 3 are entangled across two octaves of frequency difference, and that this system is therefore a potentially important resource for any quantum processes linking resources over a large bandwidth. The injection of the coherent signal allows for a simple means of control over the entanglement properties of the system. 

\section{Conclusion}

In conclusion, the proposed system is a good candidate for novel quantum technologies which need squeezed and entangled optical states spanning a wide range of frequencies. With a single cavity input field it produces three output fields which are quadrature squeezed and different pairs of modes which are EPR steerable, with selection of the desired pairs being possible either by increasing the pump power or by injected signal. The quantum correlations of interest change depending on whether the system is being operated above or below the oscillation threshold, with good EPR steering being available in both regimes. The tripartite entanglement inequalities are only violated above threshold, where the lowest frequency mode develops a non-zero mean amplitude. An injected signal at the lowest frequency removes the threshold altogether and can provide either symmetric EPR steering across one octave or asymmetric EPR steering across two octaves. The flexibility and easy controllability of this system make it an attractive candidate for experimental investigation and future technological use.

\acknowledgments

J.Y. Li was supported by the National Natural Science Foundation of China
(Grant No. 11704287). 




\begin{thebibliography}{99}


\bibitem{Armstrong} {J.A. Armstrong, N. Bloembergen, J. Ducuing, and P.S.
Pershan, Phys. Rev.\ \textbf{127}, 1918 (1962).}
%
\bibitem{Giordmaine}{J.A. Giordmaine and R.C. Miller, \prl {\bf 14}, 973 (1965).}
%
\bibitem{Nassau}{K. Nassau and H.J. Levinstein, Appl. Phys. Lett. {\bf 7}, 69 (1965).}
%
\bibitem{Yariv}{A. Yariv and W.H. Louisell, IEEE J. Quantum Elect. QE{\bf 2}, 418 (1966).}
%
\bibitem{RMPMargaret}{M.D. Reid, P.D. Drummond, W.P. Bowen, E.G. Cavalcanti, P.K. Lam, H.A. Bachor, U.L. Andersen, and G. Leuchs, \rmp {\bf 81}, 1727 (2009).}
%
\bibitem{SHGPereira}{S.F. Pereira, Min Xiao, H.J. Kimble, and J.L. Hall, \pra {\bf 38}, 4931 (1988).}
%
\bibitem{sumdiff}{M.K. Olsen and R.J. Horowicz, \oc {\bf 168}, 135 (1999).}
%
\bibitem{PingKoy}{N.B. Grosse, S. Assad, M. Mehmet, R. Schnabel, T. Symul, and P.K. Lam, \prl {\bf 100}, 243601 (2008).}
%
\bibitem{4HG}{M.K. Olsen, arXiv:1707.02537.}
%
\bibitem{4HGtri}{M.K. Olsen, \oc {\em in press}, https://doi.org/10.1016/j.optcom.2017.09.090}
%
\bibitem{multiplex} {C. Baune, J. Gniesmer, S. Kocsis, C.E. Vollmer, P. Zell,
J. Fiurasek, and R. Schnabel,  \pra {\bf 93}, 010302 (2016).}
%
\bibitem{Hammerer}{K. Hammerer, A.S. S\o rensen, and E.S. Polzik, \rmp {\bf 82}, 1041 (2010).}
%
\bibitem{Julsgaard}{B. Julsgaard, J. Sherson, I. Cirac, L. Fiur\'a\u{s}ek, and E.S. Polzik, Nature {\bf 432}, 482 (2004).}
%
\bibitem{P+} {P.D. Drummond and C.W. Gardiner, \jpa {\bf 13}, 2353
(1980).}
%
\bibitem{mjc} {C.W. Gardiner and M.J. Collett, \pra {\bf 31}, 3761 (1985).}
%
\bibitem{Zhu}{S. Zhu, Y. Zhu, and N. Ming, Science {\bf 278}, 843 (1997).}
%
\bibitem{Liz} {E. Marcellina, J .F.Corney, and M. K. Olsen \oc {\bf 309}, 9
(2013).}
%
\bibitem{Granja}{M.K. Olsen, S.C.G. Granja, and R.J. Horowicz, \oc {\bf 165}, 293 (1999).}
%
\bibitem{DFW} {D.F. Walls and G.J. Milburn, \emph{Quantum Optics}
(Springer, Berlin, 1995).}
%
\bibitem{SMCrispin} {C.W. Gardiner, \emph{Handbook of Stochastic Methods\/},
(Springer, Berlin, 1985).}
%
\bibitem{DMW}{P.D. Drummond, K.J. McNeil, and D.F. Walls, Optica Acta {\bf 28}, 211 (1981).}
%
\bibitem{Arabe}{S. Chaturvedi, K. Dechoum, and P.D. Drummond, \pra {\bf 65}, 033805 (2002).}
%
\bibitem{pulse}{K.J. McNeil, P.D. Drummond, and D.F. Walls, \oc {\bf 27}, 292 (1978).}
%
\bibitem{Bache}{M. Bache, P. Lodahl, A.V. Mamaev, M. Marcus, and M. Saffman, \pra {\bf 65}, 033811 (2002).}
%
\bibitem{3HG}{M.K. Olsen, arXiv:1706.05174.}
%
\bibitem{EPR} {A. Einstein, B. Podolsky, and N. Rosen, \pr {\bf 47}, 777 (1937).}
%
\bibitem{Erwin} {E. Schr\"odinger, Proc. Cam. Philos. Soc. {\bf 31}, 555
(1935).}
%
\bibitem{Jonesteer} {S.J. Jones, H.M. Wiseman, and A.C. Doherty, \pra {\bf
76}, 052116 (2007).}
%
\bibitem{EPRMDR} {M.D. Reid, \pra {\bf 40}, 913 (1989).}
%
\bibitem{ZYOu}{Z.Y. Ou, S.F. Pereira, H.J. Kimble, and K.C. Peng, \prl {\bf 68}, 3663 (1992).}
%
\bibitem{Wiseman}{H.M. Wiseman, S.J. Jones, and A.C. Doherty, \prl {\bf 98}, 140402 (2007).}
%
\bibitem{SFG}{M.K. Olsen and A.S. Bradley, \pra {\bf 77}, 023813 (2008).}
%
\bibitem{sapatona}{S.L.W Midgley, A.J. Ferris, and M.K. Olsen, \pra {\bf 81}, 022101 (2010).}
%
\bibitem{Handchen}{V. H\"andchen, T. Eberle, S. Steinlechner, A. Samblowski, T. Franz, R.F. Werner and R. Schnabel, Nat. Photon. {\bf 6}, 596 (2012).}
%
\bibitem{Bowles}{J. Bowles, T. V\'ertesi, M.T. Quintino, and N. Brunner, \prl {\bf 112}, 200402 (2014).}
%
\bibitem{SHGEPR}{M.K. Olsen, \pra {\bf 88}, 051802 (2013).}
%
\bibitem{vLF}{P. van Loock and A. Furusawa, \pra {\bf 67},  (2003) 052315.}
%
\bibitem{AxMuzzJPB}{M.K. Olsen and A.S. Bradley, \jpb {\bf 39}, 127 (2005).}
%
\bibitem{AxMuzz}{M.K. Olsen and A.S. Bradley, \pra {\bf 74} 063809 (2006).}
%
\bibitem{Teh&Reid}{R.Y. Teh and M.D. Reid, \pra {\bf 90}, 062337 (2014).}
%
\bibitem{Wang} {M. Wang, Y. Xiang, Q. He, Q. Gong, \pra {\bf 91}, 012112 (2015).}
%
\bibitem{OBR} {M.K. Olsen, A.S. Bradley, M.D. Reid, \jpb {\bf 39}, 2515 (2006).}
%
\bibitem{HeReid}{Q.Y. He and M.D. Reid, \prl {\bf 111}, 250403 (2013).}
%
\bibitem{Bjorkholm}{J.E. Bjorkholm and H.G. Danielmeyer,
Appl. Phys. Lett. {\bf 15}, 171 (1969).}
%
\bibitem{Haub}{J.G. Haub, M.J. Johnson, B.J. Orr, and R. Wallenstein, Appl. Phys. Lett. {\bf 58}, 1718 (1991).}
%
\bibitem{Hovde}{ D.C. Hovde, J.H. Timmermans, G. Scoles, and K.K.
Lehmann, \oc {\bf 86}, 294 (1991).}
%
\bibitem{Plusquellic}{D.F. Plusquellic, O. Votava, and D.J. Nesbitt, Appl. Optics {\bf 35}, 1464 (1996). }
%
\bibitem{kaled}{M.K. Olsen, K. Dechoum, and L.I. Plimak, \oc {\bf 223}, 123 (2003).}
%
\bibitem{signal} {M.K. Olsen, \prl {\bf 119}, 160501 (2017).}

\end{thebibliography}
\end{document}